\documentclass[10pt]{article}
\usepackage{amstext,amsfonts,amsbsy,eucal,amssymb, amsthm}
\usepackage{color}
\usepackage[dvips]{graphicx}

\begin{document}
\begin{center}
{\bf \large On various integrable discretisations of a general two component  Volterra system }
\end{center}
\begin{center}
Corina N. Babalic$^{\dagger}$, A. S. Carstea$^*${\footnote{Corresponding author:carstea@gmail.com}}
\end{center}
\begin{center}
$^\dagger${\it Dept. of Physics, University of Craiova, Romania\\
$^*$National Institute of Physics and Nuclear Engineering, Dept. of Theoretical Physics, Atomistilor 407, 077125, Magurele, Bucharest, Romania\\ }  
\end{center}
\begin{abstract}
We present two integrable discretisations of a general differential-difference bicomponent Volterra system. 
The results are obtained by discretising directly the corresponding Hirota bilinear equations in two different ways. Multisoliton solutions are presented together with a new discrete form of Lotka-Volterra equation obtained by an alternative bilinearisation.

\end{abstract}

\section{Introduction}
One of the main difficulty in the topic of integrable systems is obtaining an integrable discretisation of a given partial differential or differential-difference integrable system. Applying integrability criteria like complexity growth \cite{sebastien} or singularity confinement \cite{alfred} is not always quite easy because the discrete lattice equations have, in general, complicated forms. However, is much more convenient to start with some general lattice equation and impose some simpler (but more restrictive) integrability requirements (like, for instance, cube consistency \cite{suris} which leads immediately to the discrete variant of  ``zero curvature representation'').

Among the methods of finding integrable discretisations, one of the powerful one is the Hirota bilinear method. The idea is quite simple. First, the integrable differential or differential-difference integrable system has to be correctly bilinearised (in the sense of allowing construction of general multisoliton solution).  After that, in a first step, one has to replace differential Hirota operators with discrete ones preserving gauge invariance. Of course the resulting bilinear fully discrete system is not necesarily integrable, so in the second step, the multisoliton solution must be found \cite{side3}. If this exists then the discrete bilinear system is integrable and, in the final step which is rather complicated, the nonlinear form has to be recovered. Here, it is possible to introduce an auxiliary function, as Hirota have shown in \cite{hirota2000}. We have to clarify here the concept of integrability. Usually, the integrability for a given partial differential or partial discrete equation, is relying on the existence of an infinite number of independent integrals in involution  that usually can be computed from the Lax pairs. Accordingly, if the equation can be written as a compatibility condition of two nontrivial linear operators (i.e. Lax pairs) then it is automatically completely integrable. However, the alternative formulation given by Hirota bilinear form, is also used in proving integrability by requiring the existence of general multisoliton solution. Here the word {\it general}  means the solution describing multiple collisions of an {\it  arbitrary} number of solitons having {\it arbitrary} parameters and phases and also for {\it all}  branches of dispersion relations. Any constraint in the form of multisoliton solution breaks complete integrability.  The integrals of motion are related to the soliton parameters which, because of elastic collisions, remain unchanged (and they are also related to the spectrum of Lax operator  invariant as well since the equation itself is an isospectral deformation). Construction of the  multisoliton solution is quite difficult, but it has been observed that once three general soliton solution is constructed then it can be proved by induction that the general N-soliton can be constructed as well. The existence of three-soliton solution has been used in classification of completely integrable bilinear equations \cite{hietarinta-mkdv}, \cite{hietarinta-sg} as an integrability criterion.   

In  this paper we are going to give two integrable discretisations of a general two component bidirectional Volterra system (in the sense that the dispersion relation has two branches). This system is rather old. It has been formulated for the first time by Hirota and Satsuma \cite{hirsat} but with constraints on parameters. Then,  various solutions (rational, white and dark solitonic) have been obtained \cite{fane}, \cite{narita}, \cite{chow}. Curious enough, the fully discretisation has not been studied so far, although the initial system can be formulated as a coupled differential-difference focusing or defocusing mKdV system with non-zero boundary conditions.  In this paper we present two fully discrete bilinear forms, show the structure of $N$-soliton solution and then recover the nonlinear form. For the second discretisation we extend the approach of Hirota presented in \cite{hirota2000} and use two auxiliary functions. An important fact is that the multisoliton solution in the case of second discretisation has practically the same phase factors and interaction terms as in the differential-difference one. The same fact has been observed by Hirota and Tsujimoto in \cite{hirota2000},\cite{hir2002}, \cite{tsuji} for other cases. They have shown that for many examples including lattice mKdV, lattice NLS, lattice coupled mKdV, the structure of soliton solution remains the same as in the case of differential-difference analogs (although, it is true that all their examples support only unidirectional solitons). We think that this fact happens in many other cases and as an other example we construct a new completely integrable discretisation of Lotka-Volterra equation.

\section{Discretisation of general Volterra system}

The differential-difference system under consideration is:

\begin{eqnarray}
\dot Q_n=(c_0+c_1Q_n+c_2Q_n^2)(R_{n+1}-R_{n-1})\\
\dot R_n=(c_0+c_1R_n+c_2R_n^2)(Q_{n+1}-Q_{n-1})\nonumber
\end{eqnarray}
where $c_0, c_1, c_2$ are arbitrary constants and the dot means derivative with respect to time.
\noindent In \cite{fane} we have shown that following the simple scalings and translations,
$$u_n=\frac{2c_2}{\sqrt{4c_0c_2-c_1^2}}(Q_n+\frac{c_1}{2c_2}), \quad v_n=\frac{2c_2}{\sqrt{4c_0c_2-c_1^2}}(R_n+\frac{c_1}{2c_2})$$
we can cast the system in the following form:
\begin{eqnarray}
\dot u_n=(1+u_n^2)(v_{n+1}-v_{n-1})\\
\dot v_n=(1+v_n^2)(u_{n+1}-u_{n-1})\nonumber
\end{eqnarray}
where $u_n, v_n\to \alpha$ as $n\to \pm\infty$ and $\alpha=c_1(4c_0c_2-c_1^2)^{-1/2}$. These transformations are valid only in the case of nonzero $c_2$. Of course, one must be careful about the square root appearing in the definition of $\alpha$. In case of negative argument the system is changed in a defocusing form:
\begin{eqnarray}
\dot u_n=(1-u_n^2)(v_{n+1}-v_{n-1})\\
\dot v_n=(1-v_n^2)(u_{n+1}-u_{n-1})\nonumber
\end{eqnarray}
From the point of view of integrability this is not crucial since an overall imaginary unit factor for $u_n, v_n$ will change the system one into the other. But for the real soliton dynamics (which we are not discussing here)the solutions of the above systems are behaving completely different (only the unicomponent case has been analysed quite recently in \cite{chow}).  

For the system (2) the Hirota bilinear form is given by:
\begin{eqnarray}
\dot f_n g_n-f_n \dot g_n=(1+\alpha^2)(F_{n+1}G_{n-1}-G_{n+1}F_{n-1})\\
\dot F_n G_n-F_n \dot G_n=(1+\alpha^2)(f_{n+1} g_{n-1}-g_{n+1} f_{n-1})\\
(1+i\alpha)F_{n+1} G_{n-1}+(1-i\alpha)F_{n-1}G_{n+1}=2f_ng_n\\
(1+i\alpha)f_{n+1} g_{n-1}+(1-i\alpha)f_{n-1} g_{n+1}=2F_nG_n
\end{eqnarray}
where the nonlinear substitutions are:
$$u_n=\alpha-\frac{i}{2}\frac{\partial}{\partial t}\ln{\frac{f_n(t)}{g_n(t)}}, \quad  v_n=\alpha-\frac{i}{2}\frac{\partial}{\partial t}\ln{\frac{F_n(t)}{G_n(t)}}$$

The bilinear system (4)-(7) is completely integrable \cite{fane} and has the following $N$-soliton solution: 
%(we use here a more convenient notation which replace the usual exponential $\exp(k n+\omega t)$ with $p^n e^{\omega t}$, where $p_i=e^{k_i}$ and $\epsilon_i=\pm 1$ is the propagation direction of the soliton $i$):

$$f_n=\sum_{\mu_1...\mu_N\in\{0,1\}}\left(\prod_{i=1}^{N}(\beta_i p_i^n e^{\omega_i t})^{\mu_i}\prod_{i<j}^{N}A_{ij}^{\mu_i\mu_j}\right)$$
$$F_n=\sum_{\mu_1...\mu_N\in\{0,1\}}\left(\prod_{i=1}^{N}(\beta_i' p_i^n e^{\omega_i t})^{\mu_i}\prod_{i<j}^{N}A_{ij}^{\mu_i\mu_j}\right)$$
$$g_n=\sum_{\mu_1...\mu_N\in\{0,1\}}\left(\prod_{i=1}^{N}(\gamma_i p_i^n e^{\omega_i t})^{\mu_i}\prod_{i<j}^{N}A_{ij}^{\mu_i\mu_j}\right)$$
$$G_n=\sum_{\mu_1...\mu_N\in\{0,1\}}\left(\prod_{i=1}^{N}(\gamma_i' p_i^n e^{\omega_i t})^{\mu_i}\prod_{i<j}^{N}A_{ij}^{\mu_i\mu_j}\right)$$
where the dispersion relation and phase factors are:
$$\omega_i=\epsilon_i(1+\alpha^2)\frac{1-p_i^2}{p_i}$$
$$\beta_i=\frac{i\alpha(-1+\frac{\epsilon_i}{2}(p_i+p_i^{-1}))}{(p_i-p_i^{-1})}-\frac{\epsilon_i}{2},\quad \gamma_i=\frac{i\alpha(-1+\frac{\epsilon_i}{2}(p_i+p_i^{-1}))}{(p_i-p_i^{-1})}+\frac{\epsilon_i}{2}$$
$$\beta_i'=\frac{i\alpha\epsilon_i(1-\frac{\epsilon_i}{2}(p_i+p_i^{-1}))}{(p_i-p_i^{-1})}+\frac{1}{2},\quad \gamma_i'=\frac{i\alpha\epsilon_i(1-\frac{\epsilon_i}{2}(p_i+p_i^{-1}))}{(p_i-p_i^{-1})}-\frac{1}{2}$$
$$A_{ij}=\left(\frac{\epsilon_i p_i-\epsilon_j p_j}{1-\epsilon_i\epsilon_jp_ip_j}\right)^2$$

In order to construct an integrable discretisation we replace time derivatives in (4) and (5) with finite differences ($t\to m$) 
$$\dot f_n\to\frac{1}{\delta}(f(n,m+\delta)-f(n,m))$$
and impose the invariance of the resulting bilinear equation with respect to multiplication with $\exp(\mu n+\nu m)$ for any $\mu, \nu$ (bilinear gauge invariance). In this first discretisation we discretise {\it also} the equations (6) and (7) by assuming a gauge-invariant shift in the time variable.

The fully discrete gauge invariant bilinear equations are given by:

\begin{eqnarray}
\tilde f_n g_n-f_n \tilde g_n=\delta(1+\alpha^2)(\tilde F_{n+1} G_{n-1}-\tilde G_{n+1} F_{n-1})\\
\tilde F_n G_n-F_n \tilde G_n=\delta(1+\alpha^2)(\tilde f_{n+1} g_{n-1}-\tilde g_{n+1} f_{n-1})\\
(1+i\alpha)\tilde F_{n+1} G_{n-1}+(1-i\alpha)F_{n-1}\tilde  G_{n+1}=\tilde f_n g_n+f_n \tilde g_n\\
(1+i\alpha)\tilde f_{n+1} g_{n-1}+(1-i\alpha)f_{n-1}\tilde  g_{n+1}=\tilde F_n G_n+F_n \tilde G_n\
\end{eqnarray}
where $\tilde f_n=f(n,m+\delta)$ etc. 
In order to check the integrability one has to compute the general $N$-soliton solution of the above system. With the aid of a symbolic computation software like MATHEMATICA one can easily find 3-soliton solution. It can be extended to N-soliton solution in the form (the proof is given in the Appendix):

\begin{eqnarray}
f_n=\sum_{\mu_1...\mu_N\in\{0,1\}}\left(\prod_{i=1}^{N}(a_i p_i^n q_i^{m\delta})^{\mu_i}\prod_{i<j}^{N}M_{ij}^{\mu_i\mu_j}\right)\\
F_n=\sum_{\mu_1...\mu_N\in\{0,1\}}\left(\prod_{i=1}^{N}(A_i p_i^n q_i^{m\delta})^{\mu_i}\prod_{i<j}^{N}M_{ij}^{\mu_i\mu_j}\right)\\
g_n=\sum_{\mu_1...\mu_N\in\{0,1\}}\left(\prod_{i=1}^{N}(b_i p_i^n q_i^{m\delta})^{\mu_i}\prod_{i<j}^{N}M_{ij}^{\mu_i\mu_j}\right)\\
G_n=\sum_{\mu_1...\mu_N\in\{0,1\}}\left(\prod_{i=1}^{N}(B_i p_i^n q_i^{m\delta})^{\mu_i}\prod_{i<j}^{N}M_{ij}^{\mu_i\mu_j}\right)
\end{eqnarray}
where the dispersion relation and phase factors are given by:
\begin{equation}
q_i=\left(\frac{p_i+\delta\epsilon_i(1+\alpha^2)}{p_i+p_i^2 \epsilon_i\delta(1+\alpha^2}\right)^{1/\delta}
\end{equation}
$$a_i=\frac{i\alpha(-1+\frac{\epsilon_i}{2}(p_i+p_i^{-1}))}{(p_i-p_i^{-1})(1+\delta+\delta\alpha^2)}-\frac{\epsilon_i}{2},\quad b_i=\frac{i\alpha(-1+\frac{\epsilon_i}{2}(p_i+p_i^{-1}))}{(p_i-p_i^{-1})(1+\delta+\delta\alpha^2)}+\frac{\epsilon_i}{2}$$
\begin{equation}
A_i=\frac{i\alpha\epsilon_i(1-\frac{\epsilon_i}{2}(p_i+p_i^{-1}))}{(p_i-p_i^{-1})(1+\delta+\delta\alpha^2)}+\frac{1}{2},\quad B_i=\frac{i\alpha\epsilon_i(1-\frac{\epsilon_i}{2}(p_i+p_i^{-1}))}{(p_i-p_i^{-1})(1+\delta+\delta\alpha^2)}-\frac{1}{2}
\end{equation}
\begin{equation}
M_{ij}=\left(\frac{\epsilon_i p_i-\epsilon_j p_j}{1-\epsilon_i\epsilon_jp_ip_j}\right)^2
\end{equation}
Accordingly, our bilinear system is an integrable one. Now we can proceed to recover the nonlinear form. Dividing (8) by (10) and (9) by (11) and taking into account that $\tan(\frac{i}{2}\log(G/F))=i(G-F)/(G+F)$ for any $G$ and $F$ we obtain the following system:

$$\tan(\tilde{Q}_n-Q_n)=\frac{\delta(1+\alpha^2)\tan(\tilde{R}_{n+1}-R_{n-1})}{1+\alpha\tan(\tilde{R}_{n+1}-R_{n-1})}$$
\begin{equation}
\tan(\tilde{R}_n-R_n)=\frac{\delta(1+\alpha^2)\tan(\tilde{Q}_{n+1}-Q_{n-1})}{1+\alpha\tan(\tilde{Q}_{n+1}-Q_{n-1})}
\end{equation}
where $Q_n=\frac{i}{2}\log(f_n/g_n), R_n=\frac{i}{2}\log(F_n/G_n)$. To our knowledge this system is a new one. We do not know how it is related to other integrable discretizations of Volterra systems. In the case of $\alpha\to 0$ the classical lattice self-dual network of Hirota is obtained. 

Since the phase factors are defined up to the multiplication with the same constant factor (which we take to be the imaginary unit) then $f_n$, $g_n$ (and $F_n$, $G_n$ as well) will be complex conjugated so the physical fields $Q_n$ and $R_n$ are real functions. Here we give the 1 and 2-soliton solution (see also the figure):
$$Q_n=\frac{i}{2}\log\left(\frac{1+ia_1p_1^nq_1^{m\delta}}{1+ib_1p_1^nq_1^{m\delta}}\right),\quad R_n=\frac{i}{2}\log\left(\frac{1+iA_1p_1^nq_1^{m\delta}}{1+iB_1p_1^nq_1^{m\delta}}\right).$$

$$Q_n=\frac{i}{2}\log\left(\frac{1+ia_1p_1^nq_1^{m\delta}+ia_2p_2^nq_2^{m\delta}-a_1a_2M_{12}(p_1p_2)^n(q_1q_2)^{m\delta}}{1+ib_1p_1^nq_1^{m\delta}+ib_2p_2^nq_2^{m\delta}-b_1b_2M_{12}(p_1p_2)^n(q_1q_2)^{m\delta}}\right),$$
$$R_n=\frac{i}{2}\log\left(\frac{1+iA_1p_1^nq_1^{m\delta}+iA_2p_2^nq_2^{m\delta}-A_1A_2M_{12}(p_1p_2)^n(q_1q_2)^{m\delta}}{1+iB_1p_1^nq_1^{m\delta}+iB_2p_2^nq_2^{m\delta}-B_1B_2M_{12}(p_1p_2)^n(q_1q_2)^{m\delta}}\right).$$

\begin{figure}[ht]
\begin{center}
\includegraphics[width=6cm]{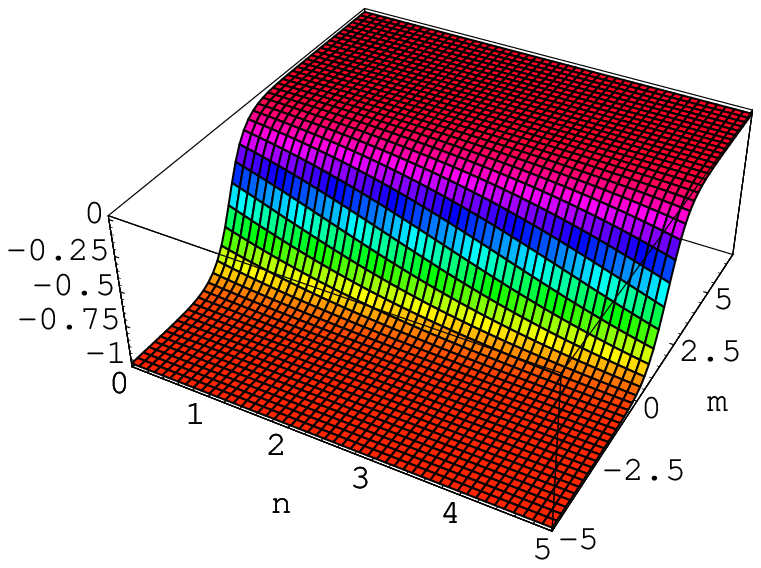}
\mbox{\raisebox{10mm}{\includegraphics[width=5cm]{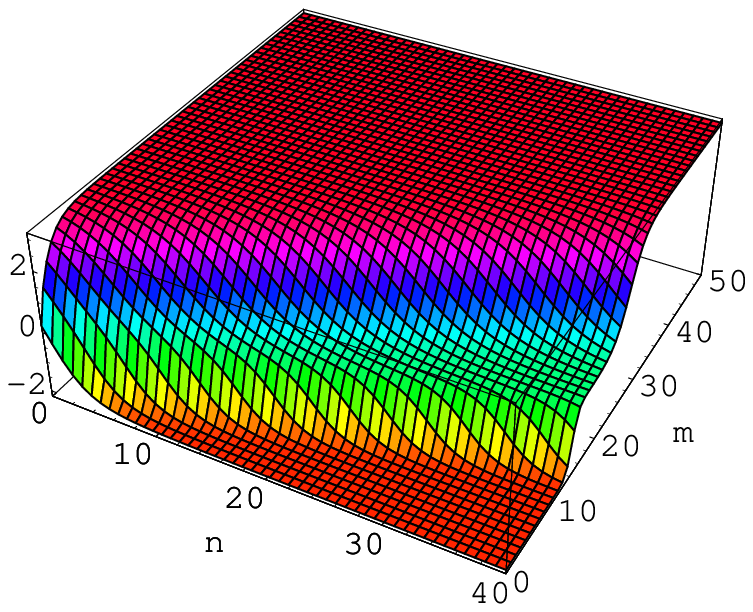}}}
\caption{One and two-soliton solution for $Q_n$ ($p_1=2, p_2=0.7, \alpha=1, \delta=2, \epsilon_1=-1, \epsilon_2=1$)}
\end{center}
\end{figure}

However one can obtain a different nonlinear form by taking $q_n=f_n/g_n$ and $r_n=F_n/G_n$, namely:
$$\tilde{q}_n=q_n\frac{z_1 \tilde{r}_{n+1}+z_2 r_{n-1}}{z_2^* \tilde{r}_{n+1}+z_1^* r_{n-1}}, \quad \tilde{r}_n=r_n\frac{z_1 \tilde{q}_{n+1}+z_2 q_{n-1}}{z_2^* \tilde{r}_{n+1}+z_1^* r_{n-1}}$$ where $z_1=1+i\alpha+\delta(1+\alpha^2), z_2=1-i\alpha-\delta(1+\alpha^2).$
In the case of $q_n=r_n$ the system becomes the Nijhoff-Capel lattice mKdV equation \cite{frank} but with complex coefficients. 
Of course a complete description imposes finding the Lax pairs, hamiltonian structure using r-matrices and so on, \cite{suris2}.

\section{Second discretisation}
We may discretise the bilinear equations (4)-(7) in a simpler way. Namely we only discretise the dispersion equations (4) and (5) which depend explicitly on time and leave (6) and (7) unmodified. We obtain:
\begin{eqnarray}
\tilde f_n g_n-f_n \tilde g_n=\delta(1+\alpha^2)(\tilde F_{n+1} G_{n-1}-\tilde G_{n+1} F_{n-1})\\
\tilde F_n G_n-F_n \tilde G_n=\delta(1+\alpha^2)(\tilde f_{n+1} g_{n-1}-\tilde g_{n+1} f_{n-1})\\
(1+i\alpha) F_{n+1} G_{n-1}+(1-i\alpha)F_{n-1} G_{n+1}=2 f_n g_n\\
(1+i\alpha) f_{n+1} g_{n-1}+(1-i\alpha)f_{n-1} g_{n+1}=2 F_n G_n
\end{eqnarray}
This bilinear system is again completely integrable. It has $N$-soliton  solution of the  same form,
\begin{eqnarray}
f_n=\sum_{\mu_1...\mu_N\in\{0,1\}}\left(\prod_{i=1}^{N}(\beta_i p_i^n q_i^{m\delta})^{\mu_i}\prod_{i<j}^{N}A_{ij}^{\mu_i\mu_j}\right)\\
F_n=\sum_{\mu_1...\mu_N\in\{0,1\}}\left(\prod_{i=1}^{N}(\beta_i' p_i^n q_i^{m\delta})^{\mu_i}\prod_{i<j}^{N}A_{ij}^{\mu_i\mu_j}\right)\\
g_n=\sum_{\mu_1...\mu_N\in\{0,1\}}\left(\prod_{i=1}^{N}(\gamma_i p_i^n q_i^{m\delta})^{\mu_i}\prod_{i<j}^{N}A_{ij}^{\mu_i\mu_j}\right)\\
G_n=\sum_{\mu_1...\mu_N\in\{0,1\}}\left(\prod_{i=1}^{N}(\gamma_i' p_i^n q_i^{m\delta})^{\mu_i}\prod_{i<j}^{N}A_{ij}^{\mu_i\mu_j}\right)
\end{eqnarray}
The dispersion relation and the phase factors are:
$$q_i=\left(\frac{p_i+\delta\epsilon_i(1+\alpha^2)}{p_i+p_i^2 \epsilon_i\delta(1+\alpha^2}\right)^{1/\delta}$$
$$\beta_i=\frac{i\alpha(-1+\frac{\epsilon_i}{2}(p_i+p_i^{-1}))}{(p_i-p_i^{-1})}-\frac{\epsilon_i}{2},\quad \gamma_i=\frac{i\alpha(-1+\frac{\epsilon_i}{2}(p_i+p_i^{-1}))}{(p_i-p_i^{-1})}+\frac{\epsilon_i}{2}$$
$$\beta_i'=\frac{i\alpha\epsilon_i(1-\frac{\epsilon_i}{2}(p_i+p_i^{-1}))}{(p_i-p_i^{-1})}+\frac{1}{2},\quad \gamma_i'=\frac{i\alpha\epsilon_i(1-\frac{\epsilon_i}{2}(p_i+p_i^{-1}))}{(p_i-p_i^{-1})}-\frac{1}{2}$$
$$A_{ij}=\left(\frac{\epsilon_i p_i-\epsilon_j p_j}{1-\epsilon_i\epsilon_jp_ip_j}\right)^2$$
The nonlinear form can be easily recovered although the system now is more complicated and it will involve two auxiliary functions $w_n$ and $v_n$. We divide (20) by $g_n\tilde{g}_n$ and (21) by $G_n\tilde{G}_n$. Calling $x_n=f_n/g_n, y_n=F_n/G_n, w_n=G_{n-1}\tilde{G}_{n+1}/g_n\tilde{g_n}, v_n=g_{n-1}\tilde{g}_{n+1}/G_n\tilde{G}_n$ we get:
$$\tilde{x}_n-x_n=\delta(1+\alpha^2)(\tilde{y}_{n+1}-y_{n-1})w_n$$
$$\tilde{y}_n-y_n=\delta(1+\alpha^2)(\tilde{x}_{n+1}-x_{n-1})v_n$$
But one can see immediately that:
$$w_{n+1}/v_n=\frac{G_n\tilde{G}_{n+2}G_n\tilde{G}_n}{g_{n+1}\tilde{g}_{n+1}g_{n-1}\tilde{g}_{n+1}}=\left(\frac{{G_{n}}^2}{g_{n+1}g_{n-1}}\right)\left(\frac{\tilde{G}_{n+2}\tilde{G}_n}{\tilde{g}_{n+1}^2}\right)$$
$$v_{n+1}/w_n=\frac{g_n\tilde{g}_{n+2}g_n\tilde{g}_n}{G_{n+1}\tilde{G}_{n+1}G_{n-1}\tilde{G}_{n+1}}=\left(\frac{{g_{n}}^2}{G_{n+1}G_{n-1}}\right)\left(\frac{\tilde{g}_{n+2}\tilde{g}_n}{\tilde{G}_{n+1}^2}\right)$$
The factors in the parantheses can be computed easily from (22) and (23) by dividing them to $G_{n-1}G_{n+1}$ and $g_{n-1}g_{n+1}$. Finally the nonlinear form of our sistem is:
$$\tilde{x}_n-x_n=\delta(1+\alpha^2)(\tilde{y}_{n+1}-y_{n-1})w_n$$
$$\tilde{y}_n-y_n=\delta(1+\alpha^2)(\tilde{x}_{n+1}-x_{n-1})v_n$$
$$w_{n+1}=v_n\frac{x_{n+1}\tilde{x}_{n+1}(1+i\alpha)+x_{n-1}\tilde{x}_{n+1}(1-i\alpha)}{y_{n}\tilde{y}_{n+2}(1+i\alpha)+y_n\tilde{y}_{n}(1-i\alpha)}$$
$$v_{n+1}=w_n\frac{y_{n+1}\tilde{y}_{n+1}(1+i\alpha)+y_{n-1}\tilde{y}_{n+1}(1-i\alpha)}{x_{n}\tilde{x}_{n+2}(1+i\alpha)+x_n\tilde{x}_{n}(1-i\alpha)}$$
We can eliminate the auxiliary functions $w_n$ and $v_n$ and we get the following higher order system:
\begin{equation}
\frac{\tilde{x}_{n+1}-x_{n+1}}{\tilde{y}_{n+2}-y_n}=\frac{\tilde{y}_n-y_n}{\tilde{x}_{n+1}-x_{n-1}}\frac{x_{n+1}\tilde{x}_{n+1}(1+i\alpha)+x_{n-1}\tilde{x}_{n+1}(1-i\alpha)}{y_{n}\tilde{y}_{n+2}(1+i\alpha)+y_n\tilde{y}_{n}(1-i\alpha)}
\end{equation}
\begin{equation}
\frac{\tilde{y}_{n+1}-y_{n+1}}{\tilde{x}_{n+2}-x_n}=\frac{\tilde{x}_n-x_n}{\tilde{y}_{n+1}-y_{n-1}}\frac{y_{n+1}\tilde{y}_{n+1}(1+i\alpha)+y_{n-1}\tilde{y}_{n+1}(1-i\alpha)}{x_{n}\tilde{x}_{n+2}(1+i\alpha)+x_n\tilde{x}_{n}(1-i\alpha)}
\end{equation}
An important remark is that we have, the {\it same} phase factors and interaction term as in the {\it differential-difference} case, namely the system (4)-(7)(and that is why we kept the same notation). Also in both equations (28), (29) the step of time discretisation $\delta$ dissapeared. So there is no trace of discretisation in the solutions, except the dispersion relation. This means that the structure of the soliton solution is the same at the level of tau functions. However, the nonlinear form is different. For instance the one and two-soliton solutions,
$$x_n=\frac{1+\beta_1p_1^nq_1^{m\delta}}{1+\gamma_1p_1^nq_1^{m\delta}},\quad y_n=\frac{1+\beta_1'p_1^nq_1^{m\delta}}{1+\gamma_1'p_1^nq_1^{m\delta}},$$
$$x_n=\frac{1+\beta_1p_1^nq_1^{m\delta}+\beta_2p_2^nq_2^{m\delta}+\beta_1\beta_2A_{12}(p_1p_2)^n(q_1q_2)^{m\delta}}{1+\gamma_1p_1^nq_1^{m\delta}+\gamma_2p_2^nq_2^{m\delta}+\gamma_1\gamma_2A_{12}(p_1p_2)^n(q_1q_2)^{m\delta}},$$
$$y_n=\frac{1+\beta_1'p_1^nq_1^{m\delta}+\beta_2'p_2^nq_2^{m\delta}+\beta_1'\beta_2'A_{12}(p_1p_2)^n(q_1q_2)^{m\delta}}{1+\gamma_1'p_1^nq_1^{m\delta}+\gamma_2'p_2^nq_2^{m\delta}+\gamma_1'\gamma_2'A_{12}(p_1p_2)^n(q_1q_2)^{m\delta}}$$

\noindent are complex functions (since the phase factors are complex). But if we define new fields $\phi_n$ and $\psi_n$ by $x_n=\exp(i\phi_n)$ and $y_n=\exp(i\psi_n)$ then the soliton solutions expressed by $\phi_n$ and $\psi_n$ are real functions and have the same shape as the ones in the first discretisation.

\noindent {\bf{Remark}}: This type of discretization involving auxiliary functions has been done also by Hirota and Tsujimoto in \cite{hirota2000},\cite{hir2002}, \cite{tsuji}. In their examples also the form and phase factors of soliton solutions remain the same. We believe that this fact happens in more general cases, although must be proved rigurously. 

Also we could have discretised the system (1) directly by taking $Q_n(t)=\Gamma_n(t)/\Phi_n(t), R_n(t)=H_n(t)/T_n(t)$ which give the bilinear system:
$$D_t \Gamma_n\cdot \Phi_n=H_{n+1}T_{n-1}-H_{n-1}T_{n+1}$$
$$D_t H_n\cdot T_n=\Gamma_{n+1}\Phi_{n-1}-\Gamma_{n-1}\Phi_{n+1}$$
$$c_0\Phi_n^2+c_1\Gamma_n\Phi_n+c_2\Gamma_n^2=T_{n-1}T_{n+1}$$
$$c_0T_n^2+c_1H_nT_n+c_2H_n^2=\Phi_{n-1}\Phi_{n+1}$$
Discretising the first two equations and after making the same steps as in the previous case we would have obtained almost a similar form:
$$\tilde{Q}_n-Q_n=\delta(R_{n+1}-R_{n-1})V_n$$
$$\tilde{R}_n-R_n=\delta(Q_{n+1}-Q_{n-1})W_n$$
$$W_{n+1}=V_n\frac{c_0+c_1\tilde{Q}_{n+1}+c_2\tilde{Q}_{n+1}^2}{c_0+c_1R_n+c_2R_n^2}$$
$$V_{n+1}=W_n\frac{c_0+c_1\tilde{R}_{n+1}+c_2\tilde{R}_{n+1}^2}{c_0+c_1Q_n+c_2Q_n^2}$$
The main drawback of this system is that the soliton solution has a very complicated form.

\section{A new form of the  Lotka-Volterra equation}
Let us give an nice and simple example related to the above construction, namely differential-difference Lotka-Volterra equation:
$$\frac{du_n}{dt}=u_n(u_{n+1}-u_{n-1})$$
Of course this equation can be seen as a particular case of the system (1) for $Q_n=R_n\equiv u_n$ and $c_0=c_2=0, c_1=1$. However because the simplified form (2) has been obtained only for $c_2\neq 0$ this equation cannot be studied as a particular case of the bilinear system (4)-(7). We consider the substitution  $u_n=G_n/F_n$ and we get the following:
$$D_t G_n\cdot F_n=(G_{n+1}F_{n-1}-G_{n-1}F_{n+1})$$
$$G_nF_n=F_{n+1}F_{n-1}$$
Now we are going to discretise only the first bilinear equation (imposing gauge-invariance in the right hand side)  and the second one will remain the same. We shall obtain ($t\to m\delta$):
\begin{equation}
\tilde{G}_nF_n-G_n\tilde{F}_{n}=\delta(\tilde{G}_{n+1}F_{n-1}-G_{n-1}\tilde{F}_{n+1})
\end{equation}
\begin{equation}
G_nF_n=F_{n+1}F_{n-1}
\end{equation}
This bilinear system is integrable because it is equivalent (if we take $G_n=f_{n-1}f_{n+2}, F_n=f_nf_{n+1}$)  with a quadrilinear equation reducible to:
\begin{equation}
\tilde{f}_nf_{n+1}+\delta f_{n-1}\tilde{f}_{n+2}-(1+\delta)f_n\tilde{f}_{n+1}=0
\end{equation}
which is the integrable bilinear form of the discrete Lotka-Volterra (with $u_n=f_{n-1}\tilde{f}_{n+2}/f_n\tilde{f}_{n+1}$) \cite{side3}
$$\tilde{u}_n=u_{n}\frac{1-\delta+\delta u_{n-1}}{1-\delta+\delta\tilde{u}_{n+1}}$$

However the nonlinear form recovered from (30) and (31) is different and we proceed as in the case of the system (28)-(29). Calling $x_n=G_n/F_n$ we get:
\begin{equation}
\tilde{x}_n-x_n=\delta(\tilde{x}_{n+1}-x_{n-1})w_n
\end{equation}
\begin{equation}
w_{n+1}=w_n\frac{\tilde{x}_{n+1}}{x_n}
\end{equation}
This is a different variant of lattice Lotka-Volterra equation. We can  solve the second linear equation in $w_n$  and we get:
$$\tilde{x}_n-x_n=\delta(\tilde{x}_{n+1}-x_{n-1})\prod_{k=-\infty}^n\frac{\tilde{x}_k}{x_{k-1}}$$
or we can eliminate $w_n$ from the first equation and we find the following nice form:
$$\frac{\tilde{x}_{n+1}-x_{n+1}}{\tilde{x}_n-x_n}\frac{\tilde{x}_{n+1}-x_{n-1}}{\tilde{x}_{n+2}-x_n}\frac{x_n}{\tilde{x}_{n+1}}=1$$
The first two factors in the left hand side look like a discrete Schwartzian derivative \cite{frank} although the expression is not a cross ratio of four points but  a cross ratio of diagonals of the two adjacent parallelograms formed by the six points. We think that this is a new equation although its symmetric form may be related in a way (unknown to us) to some well known one.

\section{Conclusions}

In this paper we have presented two integrable discretisations of a general bicomponent differential-difference Volterra system. The main procedure was discretising differential Hirota bilinear operator and then recovery of nonlinear form with the aid of some auxiliary functions. This approach may lead to higher order nonlinear equations. We apply this procedure also to the well known Lotka-Volterra equation and we found a new discrete form However we started from a different bilinearization involving two tau functions.  Relying on the fact that the structure of soliton solutions remains the same we believe that this procedure will be effective in discretising even nonintegrable equations of reaction-diffusion type (because it keeps the same structure of travelling waves).

{\bf Aknowledgements:} One of the authors (ASC) has been supported by the project PN-II-ID-PCE-2011-3-0137, Romanian Ministery of Education and Research

\section{Appendix: Proof of the N-soliton solution}
As we said in the introduction checking the existence of three soliton solution for a bilinear system is enough to prove integrability. However because the structure of the soliton solution is rather complicated we are going to sketch the proof for the N-soliton solution in the case of the first discretisation (for the second one the phase factors and bilinear equations are simpler and everything goes in the same way). We follow the same procedure as in the old papers of Hirota \cite{h1}, \cite{hs1}, \cite{hirsat}, \cite{hirkdv}. For simplicity lets call $\Delta=\delta(1+\alpha^2)$ and redefine $q_i\to q_i^{\delta}$. Plugging the N-soliton solution (12)-(15) into the first bilinear equation (8) we get
$$\sum_{\mu=0,1}\sum_{\mu'=0,1}\left[\prod_{i=1}^N a_i^{\mu_{i}}b_i^{\mu'_i}\left(\prod_{i=1}^N q_i^{\mu_i}-\prod_{i=1}^N q_i^{\mu'_i}\right)-
\Delta\prod_{i=1}^N A_i^{\mu_{i}}B_i^{\mu'_i}\left(\prod_{i=1}^N q_i^{\mu_i}p_i^{\mu_i-\mu'_i}-
\prod_{i=1}^N q_i^{\mu'_i}p_i^{\mu'_i-\mu_i}\right)\right]\times $$
\begin{equation}
\times\prod_{i<j}M_{ij}^{\mu_i\mu_j+\mu'_i\mu'_j}\prod_{i=1}^N p_i^{(\mu_i+\mu'_i)n}q_i^{(\mu_i+\mu'_i)m}=0
\end{equation}
Let the coefficient of the factor $\prod_{i=1}^{\nu_0}p_i^nq_i^m\prod_{i=\nu_0+1}^{\nu_1}p_i^{2n}q_i^{2m}$ be $F$. We have
$$F=\sum_{\mu,\mu'=0,1}c_{\mu\mu'}\left[\prod_{i=1}^{\nu_1} a_i^{\mu_{i}}b_i^{\mu'_i}\left(\prod_{i=1}^{\nu_0} q_i^{\mu_i}-\prod_{i=1}^{\nu_0} q_i^{\mu'_i}\right)-
\Delta\prod_{i=1}^{\nu_1} A_i^{\mu_{i}}B_i^{\mu'_i}\left(\prod_{i=1}^{\nu_0} q_i^{\mu_i}p_i^{\mu_i-\mu'_i}-
\prod_{i=1}^{\nu_0} q_i^{\mu'_i}p_i^{\mu'_i-\mu_i}\right)\right]\times $$
$$\times\prod_{i<j}M_{ij}^{\mu_i\mu_j+\mu'_i\mu'_j}$$
where $c_{\mu\mu'}$ implies that the summation over families of indices $\mu$ and $\mu'$ should be done under the requirements:
$$\mu_i+\mu'_i=1, i=1,...,\nu_0$$
$$\mu_i=\mu_i'=1, i=\nu_0+1,...,\nu_1$$
$$\mu_i=\mu_i'=0, i=\nu_1+1,...,N$$
Substituting the expressions of dispersion relation and phase factors (16)-(18) and introducing the multiindex $\sigma=\mu-\mu'$ we find that $F=const.\hat{F}$ where

$$\hat{F}=\sum_{\sigma=\pm 1}\{\left(\prod_{i=1}^{\nu_0}(1+\epsilon_i\Delta p_i^{\sigma_i})-\prod_{i=1}^{\nu_0}(1+\epsilon_i\Delta p_i^{-\sigma_i})\right)\prod_{i=1}^{\nu_0}\left(i\alpha(-1+\cosh(\sigma_i k_i))-\epsilon_i\sinh(\sigma_i k_i)(1+\Delta)\right)-$$
$$-\Delta\left(\prod_{i=1}^{\nu_0}(1+\epsilon_i\Delta p_i^{\sigma_i})p_i^{\sigma_i}-\prod_{i=1}^{\nu_0}(1+\epsilon_i\Delta p_i^{-\sigma_i})p_i^{-\sigma_i}\right)\prod_{i=1}^{\nu_0}\left(i\alpha\epsilon_i(1-\epsilon_i\cosh(\sigma_i k_i))+\sinh(\sigma_i k_i)(1+\Delta)\right)\}$$
$$\times\prod_{i<j}\left(\epsilon_i\epsilon_j-\cosh(\sigma_i k_i-\sigma_j k_j)\right)^2$$

\noindent Making the notation $x_i=(\epsilon_i p_i)^{1/2}$, $z=(1+\Delta+i\alpha)/2, z^*=(1+\Delta-i\alpha)/2$ the above relation becomes proportional to the following expression:
$$\hat{F}=\sum_{\sigma=\pm 1}\{\left(\prod_{j=1}^{\nu_0}(1+\Delta x_j^{2\sigma_j})-\prod_{j=1}^{\nu_0}(1+\Delta x_j^{-2\sigma_j})\right)\prod_{j=1}^{\nu_0}(-i\alpha-z^* x_j^{2\sigma_j}+zx_j^{-2\sigma_j})-$$
$$-\Delta\left(\prod_{j=1}^{\nu_0}(1+\Delta x_j^{2\sigma_j})x_j^{2\sigma_j}-\prod_{j=1}^{\nu_0}(1+\Delta x_j^{-2\sigma_j})x_j^{-2\sigma_j}\right)\prod_{j=1}^{\nu_0}(i\alpha+z^* x_j^{2\sigma_j}-zx_j^{-2\sigma_j})\}$$
$$\times\prod_{i<j}(x_i^{\sigma_i}x_j^{-\sigma_j}-x_i^{-\sigma_i}x_j^{\sigma_j})^2$$
Now in order to show that (35) holds we have to prove that $\hat{F}=0$ for any $n=1,...,N$. We shall prove this by induction. 
\noindent First, it is easily seen that this expression has the following properties:

\noindent i)$\hat{F}(x_1,...,x_n)$ is a symmetric and even function of $x_1,...,x_n$ and invariant under the transformation $x_j\to1/x_j$ for any $j$.

\noindent ii) $\hat{F}(x_1,...,x_n)|_{x_1=1}=0$

\noindent iii)$\hat{F}(x_1,...,x_n)|_{x_1=x_2}=2(1+\Delta x_1^2)(1+\Delta x_1^{-2})(-i\alpha-z^*x_1^2+zx_1^{-2})(-i\alpha-z^*x_1^{-2}+zx_1^{2})\hat{F}(x_3,...,x_n).$

\noindent Now we assume that $\hat{F}=0$ for $n-1$ and $n-2$. From ii) and iii) $\hat{F}(x_1,...,x_n)|_{x_i=1}=0$ and $\hat{F}(x_1,...,x_n)|_{x_i=x_j}=0$ for any $i,j$. Then, based on the property i), Hirota proved \cite{h1}, \cite{hs1}, \cite{hirkdv} that the expression of $\hat{F}$ can be factored by a function:
$$\frac{\prod_{i=1}^n(x_i^2-1)^2\prod_{i<j}^{(n)}(x_i^2-x_j^2)^2(x_i^2x_j^2-1)^2}{\prod_{i=1}^nx_i^2\prod_{i<j}^{(n)}x_i^4x_j^4}$$
which has $2n+2n(n-1)$ zeros of order $2$. Accordingly $\hat{F}$ would have at least $4n^2$ polynomials in the numerator. On the other hand the function:
$$ \hat{F}(x_1,...,x_n)\prod_{i=1}^nx_i^2\prod_{i<j}^{(n)}x_i^4x_j^4$$ are polynomials of degree $2n^2+6n$ at most. Hence $\hat{F}=0$ for any $n$. 
In the same way one can prove that the N-soliton solution obeys the other bilinear equations.

For instance in the case of the third bilinear equation (10) introducing the expressing of N-soliton solution (12)-(15) we obtain
$$\sum_{\mu,\mu'=0,1}\left[\prod_{j=1}^N a_j^{\mu_{j}}b_j^{\mu'_j}\left(\prod_{j=1}^N q_j^{\mu_i}+\prod_{j=1}^N q_j^{\mu'_j}\right)-
\prod_{j=1}^N A_j^{\mu_{j}}B_j^{\mu'_j}\left(\prod_{j=1}^N(1+i\alpha) q_j^{\mu_j}p_j^{\mu_j-\mu'_j}+
\prod_{j=1}^N(1-i\alpha) q_j^{\mu'_j}p_j^{\mu'_j-\mu_j}\right)\right]\times $$
\begin{equation}
\times\prod_{i<j}M_{ij}^{\mu_i\mu_j+\mu'_i\mu'_j}\prod_{i=1}^N p_i^{(\mu_i+\mu'_i)n}q_i^{(\mu_i+\mu'_i)m}=0
\end{equation}
From this expression we obtain by the same steps as above that the coefficient of the factor $\prod_{i=1}^{\nu_0}p_i^nq_i^m\prod_{i=\nu_0+1}^{\nu_1}p_i^{2n}q_i^{2m}$ is proportional to the following expression: 
$$\hat{K}=\sum_{\sigma=\pm 1}\{\left(\prod_{j=1}^{\nu_0}(1+\Delta x_j^{2\sigma_j})+\prod_{j=1}^{\nu_0}(1+\Delta x_j^{-2\sigma_j})\right)\prod_{j=1}^{\nu_0}(-i\alpha-z^* x_j^{2\sigma_j}+zx_j^{-2\sigma_j})-$$
$$-\left(\prod_{j=1}^{\nu_0}(1+i\alpha)(1+\Delta x_j^{2\sigma_j})x_j^{2\sigma_j}+\prod_{j=1}^{\nu_0}(1-i\alpha)(1+\Delta x_j^{-2\sigma_j})x_j^{-2\sigma_j}\right)\prod_{j=1}^{\nu_0}(i\alpha+z^* x_j^{2\sigma_j}-zx_j^{-2\sigma_j})\}$$
$$\times\prod_{i<j}(x_i^{\sigma_i}x_j^{-\sigma_j}-x_i^{-\sigma_i}x_j^{\sigma_j})^2$$
Immediately one can see that $\hat{K}$ is an invariant, symmetric and even function satisfying the same properties $\hat{K}(x_1,...,x_n)|_{x_1=1}=0$, $\hat{K}(x_1,...,x_n)|_{x_1=x_2}=2(1+\Delta x_1^2)(1+\Delta x_1^{-2})(-i\alpha-z^*x_1^2+zx_1^{-2})(-i\alpha-z^*x_1^{-2}+zx_1^{2})\hat{K}(x_3,...,x_n)$, so the induction method can be applied in the same way.

\end{document}